\documentstyle[aps,prd,eqsecnum]{revtex}
\input psfig.sty

\def\be{\begin{equation}}
\def\ee{\end{equation}}
\def\ba{\begin{eqnarray}}
\def\ea{\end{eqnarray}}

\begin{document}

\draft

\title{The twin paradox in compact spaces}

\author{John D. Barrow${}^{(1)}$ and Janna Levin${}^{(1),(2)}$}
\address{${}^{(1)}$ DAMTP, Cambridge University,
Wilberforce Rd., Cambridge CB3 0WA }
\address{${}^{(2)}$ The Blackett Laboratory, Imperial College of
Science, Technology \& Medicine, South Kensington, London SW7 2BZ}

\twocolumn[\hsize\textwidth\columnwidth\hsize\csname
           @twocolumnfalse\endcsname

\maketitle
\widetext

\begin{abstract}
Twins travelling at constant relative velocity will each see the
other's time dilate leading to the apparent paradox that each twin
believes the other ages more slowly.  In a finite space, the twins
can both be on inertial, periodic orbits so that they have the
opportunity to compare their ages when their paths cross.
As we show, they will agree on their respective ages and avoid
the paradox.  The resolution relies on the selection of a preferred
frame singled out by the topology of the space.
\end{abstract}

\medskip
\noindent{}
\medskip
]

\narrowtext
 
\setcounter{section}{1}

The twin paradox in special relativity has a simple formulation and 
resolution in infinite flat space.  One twin remains on Earth while 
the other moves at constant velocity in a spaceship to a distant planet,
turns around and returns home to Earth.  Each twin believes the other's
clock runs slower and so the paradox arises that each believes the other
should be younger at their reunion.  The paradox is resolved
since the twin in the spaceship had to slow down, stop at the distant planet,
turn around, and accelerate to constant velocity before returning
to Earth.  Therefore the travelling twin was not always in an inertial frame
and special relativity is not contradicted by the realization that 
the twin who left Earth is younger than her sibling at the time of their
reunion.

In a compact space, the paradox is more complicated.  If the travelling twin 
is on a periodic orbit, she
can remain in an inertial frame for all time as she travels around the 
compact space, never stopping or turning.
Since both twins are inertial, both should see the other
suffer a time dilation.  The paradox again
arises that both will believe the other to be
younger when the twin in the rocket flies by.  The
twin paradox can be resolved in compact space and we will show that the twin
in the rocket is in fact younger than her sibling after a complete transit
around the compact space.
The resolution hinges on the existence of a preferred frame introduced
by the topology, one consequence of which is the inability of the 
twin in the rocket to synchronize her clocks 
\cite{{peters},{lh}}.  While other authors have come to similar conclusions
\cite{{peters},{lh},{weeks}}, the present discussion offers a 
completely general solution
and does not rely on any specific topology.  We also make use of the modern
language of topology which has recently seen application in cosmology 
\cite{cqg}.

The manifold of special relativity is $R \otimes {\cal M}$ where 
$R$ represents the time direction and ${\cal M}=R^3$ is a flat $3D$
infinite space.  
The flat spacetime metric is the familiar
	\be
	ds^2=g_{\mu\nu}dx^\mu dx^\nu
	\ee
with $g^\mu_\nu={\rm diag}(-1,1,1,1)$ and
	\be
	x^\mu=\pmatrix{t\cr x \cr y \cr z }\ \ .\label{coor}
	\ee
The spacetime is invariant under the action
of the Poincar\'e group,
which contains translations, rotations, and
the Lorentz transformations representing relative
motion at constant velocity.  The isometries can be represented as 
$O(3,1)$ matrices.
We consider special relativity in a compact
$3$-manifold
${\cal M}_c=R \otimes {\cal M}/\Gamma $.  The elements $\phi\in \Gamma$
act discretely,
without fixed points,
and are a subset of the full isometry group.  
The group $\Gamma $
can be thought of as the set of instructions for compactifying the space.  All
multiconnected, flat topologies can be constructed from either a parallelepiped
or a hexagonal prism with opposite sides identified according to the
rules
given by the elements
$\phi\in \Gamma$ \cite{{wolf},{luminet},{sss},{us}}.  

\begin{figure}
\centerline{\psfig{file=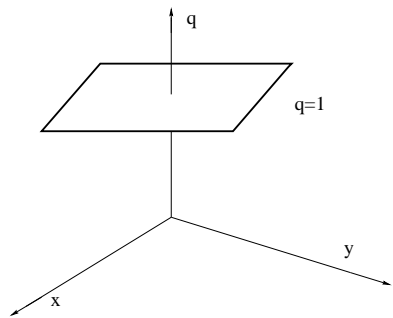,width=2.in}}
\vskip 5truept
\caption{The embedding of $(3+1)$-Minkowski space into $(4+1)$-Minkowski
space.  The $(t,z)$ directions are suppressed so that the manifold appears
as an infinite sheet fixed at $q=1$.}
\label{embed}
\end{figure}

It is 
advantageous to embed the $(3+1)$-dimensional spacetime in a 
$(4+1)$-dimensional Minkowski spacetime with the fourth spatial coordinate
fixed.  Specifically, the $(3+1)$-dimensional coordinate (\ref{coor})
is replaced with the $(4+1)$-dimensional coordinate
	\be
	x^a=\pmatrix{t\cr x \cr y \cr z \cr q}
	\ee
where $q$ is fixed at unity as in fig. \ref{embed}.  
We will let Greek indices run over $0,1,2,3$ and Latin indices run
over 
$0,1,2,3,4$.

In this coordinate system the generators can be represented as $5\times 5$
matrices. 
For instance, the generator which effects the identification of a point
$(t,x,y,z,q)$ with the point $(t,x+L_x,y,z,q)$ can be written as
\cite{jeff}
	\be
	T_x=\pmatrix{1 & 0 & 0 & 0 & 0 \cr
	0 & 1 & 0 & 0 & L_x \cr
	0 & 0 & 1 & 0 & 0 \cr
	0 & 0 & 0 & 1 & 0 \cr
	0 & 0 & 0 & 0 & 1 }\label{tx}
	\ee
so that the boundary condition can be expressed as 
$x\rightarrow T_x x$, which generalizes to 
	\be	
	x^a \rightarrow \phi^a_b x^b  \label{sbc}
	\ee
for each $\phi \in \Gamma$.
As an illustration, the hypertorus is constructed by gluing opposite
faces of the parallelepiped.  The elements of $\Gamma $ are $T_x$
of eqn (\ref{tx}) and
	\ba
	T_y=\pmatrix{1 & 0 & 0 & 0 & 0 \cr
	0 & 1 & 0 & 0 & 0 \cr
	0 & 0 & 1 & 0 & L_y \cr
	0 & 0 & 0 & 1 & 0 \cr
	0 & 0 & 0 & 0 & 1 }
,
	T_z=\pmatrix{1 & 0 & 0 & 0 & 0 \cr
	0 & 1 & 0 & 0 & 0 \cr
	0 & 0 & 1 & 0 & 0 \cr
	0 & 0 & 0 & 1 & L_z \cr
	0 & 0 & 0 & 0 & 1 } \nonumber .
	\ea
Another allowed compact topology is one that
first twists the $z$-faces through $\pi$ before 
identification.  The elements of $\Gamma$ for the twisted space
are $T_x,T_y,R_z(\pi)T_z$ with $R_z(\theta)$ the rotation matrix:
	\be
	R_z(\theta)=\pmatrix{1 & 0 & 0 & 0 & 0\cr
	0 & \cos \theta & \sin \theta & 0 & 0\cr
	0 & -\sin \theta & \cos \theta & 0 & 0\cr
 	0 & 0 & 0 & 1 & 0 \cr
 	0 & 0 & 0 & 0 & 1 \cr
	}\  \ .
	\ee
All of the multiconnected, flat topologies can be built out of a
combination of these translations and rotations.

\begin{figure}
\centerline{\psfig{file=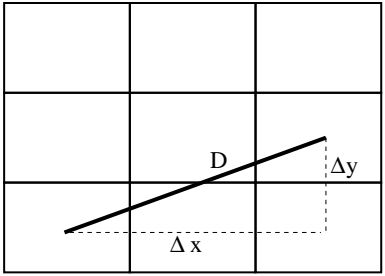,angle=-90,width=2.in}}
\vskip 5truept
\caption{
The compact hypertorus can be represented as an identified
parallelepiped.  Alternatively, the compact topology can be 
represented
by tiling flat space 
with identical copies of the fundamental 
parallelepiped.  In the tiling picture above
only the $(x,y)$ directions are shown.
A particular
periodic orbit is drawn which corresponds
to 
$x_{\rm end}=T_yT_x^2x_{\rm start}$.
}
\label{tile}
\end{figure}

Periodic orbits are of particular interest since an observer on a
periodic orbit can remain inertial.
A periodic orbit can be described by the holonomies, 
$\phi \in \Gamma$, which map the end-point of the orbit to the starting
point of the orbit.  In other words, a periodic orbit has
$x_{\rm end}=\phi x_{\rm start}$ where $\phi $ can be a composite
word
$\phi=\prod^n_i\phi_{k_i}$.  
Each word has a corresponding periodic orbit.  
For example, consider the periodic orbit of Fig. \ref{tile} in the
hypertorus.  For this orbit we have $x_{\rm end}=T_yT_x^2x_{\rm start}$.

\begin{figure}
\centerline{\psfig{file=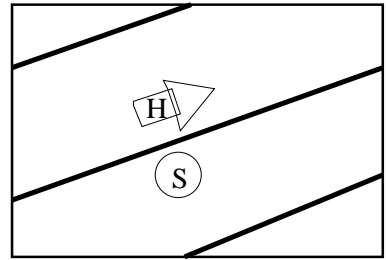,width=2.in}}
\vskip 5truept
\caption{S stays on Earth while H travels on the periodic orbit of
Fig.
\ref{tile}.
}
\label{fund}
\end{figure}

Suppose the space is compactified so that with respect to an observer S,
only spatial points are identified.  In the coordinate system at rest
with respect to S, all of the holonomies have $\phi^0_a=1$.
S's twin, H, takes a rocket ride around the compact space, travelling 
always with constant velocity, never turning, slowing or speeding up
(Fig. \ref{fund}).
A coordinate system at rest with respect to H is given by
$\bar x=\Lambda x$ with $\Lambda$ the Lorentz transformation.
In $(4+1)$ dimensions 
we can represent the most general Lorentz transformation 
as 
	\ba
	\Lambda=\pmatrix{
	\gamma & -\gamma \beta_x & -\gamma \beta_y & -\gamma \beta_z & 0 \cr
	-\gamma \beta_x & 1+{(\gamma -1)\beta_x^2\over \beta^2} &
	{(\gamma-1)\beta_x \beta_y \over \beta^2} &
	{(\gamma-1)\beta_x \beta_z \over \beta^2} & 0 \cr
	-\gamma \beta_y & 
	{(\gamma -1)\beta_x\beta_y\over \beta^2} &
	1+{(\gamma-1)\beta_y^2 \over \beta^2} &
	{(\gamma-1)\beta_y \beta_z \over \beta^2} & 0 \cr
	-\gamma \beta_z & 
	{(\gamma -1)\beta_x\beta_z\over \beta^2} &
	{(\gamma-1)\beta_y\beta_z \over \beta^2} &
	1+{(\gamma-1)\beta_z^2 \over \beta^2} & 0 \cr
	0 & 0 & 0 & 0 & 1 }\nonumber
	\ea
where the $\beta_i$ are the velocities of the 
boosts in the $(x,y,z,q)$ directions,
$\beta^2=\sum_i \beta_i^2$, and $\gamma=1/\sqrt{1-\beta^2}$.
The velocity in the $q$ direction is understood to be zero.

S sees H travel a distance $D$ before passing overhead.
During one orbital period, S's clock advances a time
	\be
	\Delta t=D/\beta
	\label{start}
	\ \ .
	\ee
However, S will believe H's clock runs slower by the factor
	\be 
	\Delta \bar t=\Delta t/\gamma
	\ee
and so expects H's clock to advance by
	\be
	\Delta \bar t=D/(\gamma \beta)
	\ \ .\label{stime}
	\ee
H is therefore younger than S when their paths coincide.

There is no paradox since H will agree
that in fact she is younger than her twin.
According to H, both space and time points have been identified.
As a result it becomes impossible for H to synchronize her clocks
\cite{peters}.
H must be on a periodic orbit to remain inertial.
Let $H$ be on a periodic orbit corresponding to the composite
word $\phi$ so
the boundary condition (\ref{sbc}) becomes 
$\bar x=\Lambda x \rightarrow \Lambda \phi x$.  The lack of synchronicity
will be given by the time component of $\Lambda(1-\phi)x$ or explicitly
	\ba
	\delta \bar t &=& (\gamma t -\gamma \beta^ix_j)
	-(\gamma t - \gamma \beta^i \phi_i^jx_j) \nonumber \\
	&=& -\gamma \beta^i(x_i-\phi_i^j x_j)  
	\ea
with $i=1,2,3,4$ and the vector $\beta^i=(\beta_x,\beta_y,\beta_z,0)$,
while $x^i=(x,y,z,q)$.
The distance travelled as measured by S is $D^2=\Delta x^a\Delta x_a$ or
	\be
	D=\sqrt{(x_i-\phi_i^j x_j) (x^i-\phi_j^ix^j) }
	\label{d}
	\ee
and
	\be
	(x_i-\phi_i^j x_j)= D\beta_i/\beta \ \ 
	\ee
so that H's clocks are out of synchronization by a factor
	\be
	\delta \bar t=-\gamma \beta D \ \ .\label{synch}
	\ee
H sees her twin S move away from her in the opposite direction only to return
after travelling a distance $\gamma D$.   With the  additional time offset 
of eqn (\ref{synch}) due to
the compact topology, H's clock must
read
	\be
	\Delta \bar t=\gamma D/\beta -\gamma \beta D=D/(\gamma \beta)
	\label{fin}
	\ee
in agreement with (\ref{stime}).  Both twins agree that H is younger
than S \cite{foot}.

Notice that ultimately the age difference between the twins is independent of
topology except through the distance $D$.  
For the orbit of Fig. \ref{tile}, for instance, $\phi=T_yT_x^2$ and
eqn. (\ref{d}) gives $D=\sqrt{2 L_x^2+L_y^2}$.

The previous example can be recast in a more physical, less abstract
discussion.  What the above formalism shows is that only one
reference frame can be at rest
with respect to the compact spatial sections.  
All other inertial observers in relative motion live in a universe
where both space and time points are identified.
In the example given 
around eqns. (\ref{start})-(\ref{fin}),
twin $S$ is at rest in a flat torus and
$H$ moves inertially
along a periodic orbit.  Suppose $H$ is initially unaware that spacetime is
compact.  In order to properly perform any experiments, $H$ has to equip
her reference frame with a full system of rulers and clocks.
She can set up a system of observers
one by one trying to synchronize their clocks by exchanging information
with a lightbeam.   
Somewhere along the way however $H$ will receive her own message telling
her to reset her clock by the amount $\gamma \beta D$.  She will be out
of synch with her
own attempts to synchronize.  That is,
observers at the same spacetime point
can have clocks that read different times.  
$H$ will know that any measurements made in this frame
are ambiguous by the time shift.

The twin paradox shows that the compact topology identifies a 
preferred frame, namely the frame in which the length along a given
side is shortest, a point emphasized in Refs. 
\cite{peters} (see also Refs. \cite{lh}).
To generalize the effect to curved space,
$\Lambda$ can be replaced by an appropriate diffeomorphism and 
the spacetime topology generalizes to
${\cal M}_c=R\otimes {\cal M}^U/\Gamma$ where 
the universal cover, ${\cal M}^U$ is a curved, simply connected
manifold.
Multiconnected cosmologies challenge the Copernican Principle.
A compact topology selects a preferred place and a preferred time so that
{\it some } galaxy, if not our own, is at the center of the universe.
Some observers are also uniquely able to synchronize their clocks
and observe the smallest volume for the universe.

\vskip 15truept

We thank P. Ferreira, 
N.J. Cornish, W.T.Gowers, A. Kent, R. Jones, G. Starkman, 
and J. Weeks 
for discussions.
JL is grateful to the theoretical physics group at
Imperial
College for their hospitality.  JL is supported by PPARC.

\end{document}